 \documentclass[epj]{svjour}
\usepackage[dvips]{graphicx}
\newcommand{\Boltz}{\mbox{$k_{\rm B}$}}
\newcommand{\sbond}{\makebox[5pt]{
\put(0,-2){\line(0,1){9}}
\put(0,-2){\circle*{2.5}}
\put(0,7){\circle*{2.5}}}}
\newcommand{\tbond}{\makebox[5pt]{
\put(0,-2){\line(0,1){3}}
\put(0,4){\line(0,1){3}}
\put(0,-2){\circle*{2.5}}
\put(0,7){\circle*{2.5}}}}
\newcommand{\hbond}{\makebox[20pt]{
\put(-5,-1){\line(1,0){10}}
\put(-5,-1){\circle*{2.5}}
\put(5,-1){\circle*{2.5}}
\put(-5,6){\line(1,0){10}}
\put(-5,6){\circle*{2.5}}
\put(5,6){\circle*{2.5}}}}
\newcommand{\xbond}{\makebox[20pt]{
\put(-5,-1){\line(6,5){10}}
\put(-5,-1){\circle*{2.5}}
\put(4,-1){\circle*{2.5}}
\put(-5,6){\line(6,-5){10}}
\put(-5,6){\circle*{2.5}}
\put(4,6){\circle*{2.5}}}}
\begin{document}
\title{Zero temperature phase transitions in spin-ladders: phase diagram
and dynamical studies of $\rm\bf Cu_2(C_{5}H_{12}N_{2})_2Cl_4$}
\author{G. Chaboussant\inst{1}\thanks{\emph{Present address:} ISIS 
Facility, Rutherford Appleton Laboratory, Didcot, Oxon OX11 0QX, United 
Kingdom.}, M.-H.
Julien\inst{1}\thanks{\emph{Present address:} Dpt. of Physics "A. Volta"
Universita di Pavia, Via Bassi 6, I-27100 Pavia, Italy.}, Y.
Fagot-Revurat\inst{1}\thanks{\emph{Present address:}  Physikalisches 
Institut, Universit{\"a}t Stuttgart, Pfaffenwaldring 57, D-70550 
Stuttgart, Germany.}, M. 
Hanson\inst{1}, L. P. L\'evy\inst{1,2}, C.
Berthier\inst{1,3}, M. Horvati\'c\inst{1} and O. Piovesana\inst{4}}
\institute{Grenoble High Magnetic Field Laboratory, CNRS and MPI-FKF, BP
166, F-38042 Grenoble, France \and
Institut Universitaire de France and Universit\'e Joseph Fourier, BP 53,
F-38400 St Martin d'H\`eres, France \and
Laboratoire de Spectrom\'etrie Physique,  Universit\'e Joseph 
Fourier, BP 87, F-38402 St Martin d'H\`eres, France \and
Dipartimento di Chimica, Universit\`a di Perugia, I-06100 
Perugia, Italy.}
\authorrunning{G. Chaboussant et al.}
\titlerunning{Zero temperature phase transitions in spin-ladders}
\date{Received: \today }

\abstract{
In a magnetic field, spin-ladders undergo two zero-temperature phase
transitions at the critical fields $H_{c1}$ and $H_{c2}$.  An
experimental review of static and dynamical properties of spin-ladders
close to these critical points is presented.  The scaling functions,
universal to all quantum critical points in one-dimension, are extracted
from (a) the thermodynamic quantities (magnetization) and (b) the
dynamical functions (NMR relaxation).  A simple mapping of strongly
coupled spin ladders in a magnetic field on the exactly solvable XXZ
model enables to make detailed fits and gives an overall understanding
of a broad class of quantum magnets in their gapless phase (between
$H_{c1}$ and $H_{c2}$).  In this phase, the low temperature divergence 
of the NMR relaxation demonstrates its Luttinger liquid nature as well 
as the novel quantum critical regime at higher temperature.
The general behavior close these quantum critical points can be tied
to known models of quantum magnetism.
} 
\maketitle
%
\section{Introduction}
It is well known that long-range order is destroyed by quantum
fluctuations in one-dimensional antiferromagnets.  If the importance of
quantum effects is ubiquitous in one-dimension, a wide variety of ground
states can nevertheless be found in nature.  Some systems have a
continuum of low energy modes, some have an energy gap above a unique
ground state, other dimerize.  Where do these differences come from?  In
simple terms, the role of quantum effects is simply to ''connect''
different classical ground states (for example the N\'eel states
$|\uparrow, \downarrow, \uparrow, \downarrow, \ldots \rangle$ and
$|\downarrow, \uparrow, \downarrow, \uparrow, \ldots \rangle$) by
tunneling processes.  Depending of the strength of the tunneling matrix
elements, which can usually be measured by a coupling constant $g$, the
system will be more or less localized around a classical ground states.
As $g$ is varied, the system can delocalize at a critical value
$g_c$.  When the system delocalizes in spin space, the ground state
becomes a rotationally invariant singlet and in all the cases which will
be considered here, an energy gap $\Delta$ appears simultaneously in the
energy spectrum.  Many physical aspects determine the strength of
quantum fluctuations.  The integer or half-integer nature of the spin
considered modify drastically selections rules for
quantum-processes\cite{Haldane83}.  This is why integer-spin chains for
which $g$ exceed $g_c$ have an energy gap, while half-integer
spin-chains remain gapless ($g \le g_c$).  Other physical parameters
(exchange constants, applied magnetic fields) also enter in the precise
determination of the coupling strength $g$.  Systems for which the
coupling constant can be continuously varied by an experimentally
controllable parameter, such as a magnetic field, are rare.  In this
paper we review a few 1-D antiferromagnets where such zero temperature
critical points have been observed, with a particular emphasis on $\rm
Cu_2(C_{5}H_{12}N_{2})_2Cl_4$ (also known as
CuHpCl)\cite{Chaboussant97a,Chaboussant97b,Chaboussant98}, a spin-ladder
compound, where a complete set of experiments exist.

At a quantum critical point
\cite{Hertz76,Chubukov94,Sachdev94,Sachdev97,Sondhi97}, the system
switches from one ground state into another.  Specifically when $g \le
g_c$, antiferromagnetic correlation functions decay as power laws and
the system is nearly ordered.  When $g$ is increased above $g_c$, a gap
opens up and the range of antiferromagnetic correlation become finite,
of the order of $\xi_g \simeq a/|g - g_c|$ ($a$ is the lattice
constant).  In the vicinity of $g_c$, one has to go to relatively long
lengthscale, exceeding $\xi_g$ to be able to tell in which phase the
system is.  In other words, the nature of the ground state is manifest
only at long lengthscale.  At finite temperature spin-flip processes
become possible.  They cut spin-correlations off at a lengthscale
$\xi_T$ which can be estimated in the quantum-disordered phase ($g \ge
g_c$) as the mean distance between excitations.  Since their energies
are higher than the energy gap $\Delta \simeq 1/|g - g_c|$ above the 
ground
state, their density is activated.  Hence when $\Boltz T < \Delta$, the
mean distance between excitations $\xi_T$ greatly exceed $\xi_g$ and
thermal fluctuations are not really relevant.  On the other hand, when
$\Boltz T \ge \Delta$, their density is governed by the relative value
of $\Boltz T - \Delta$ compared to the bandwidth of the triplet
excitations.  When $\Delta$ is small compared to $\Boltz T$ and this
bandwidth, $\xi_g$ exceeds very rapidly $\xi_T$.  In this case, the
density of excitations are determined by $\Boltz T$ {\em alone} which
become the only relevant energy scale.  In this limit, dynamical
properties are similar to those of a simple paramagnet ($1/T_1
\rightarrow cst$) while thermodynamics quantities remain nontrivial.
This regime is (improperly) named the {\em quantum critical regime},
because most properties are determined by the single lengthscale $\xi_T$
as in ordinary phase transitions.
\begin{figure}
\centering \includegraphics[height=70mm, angle=0]{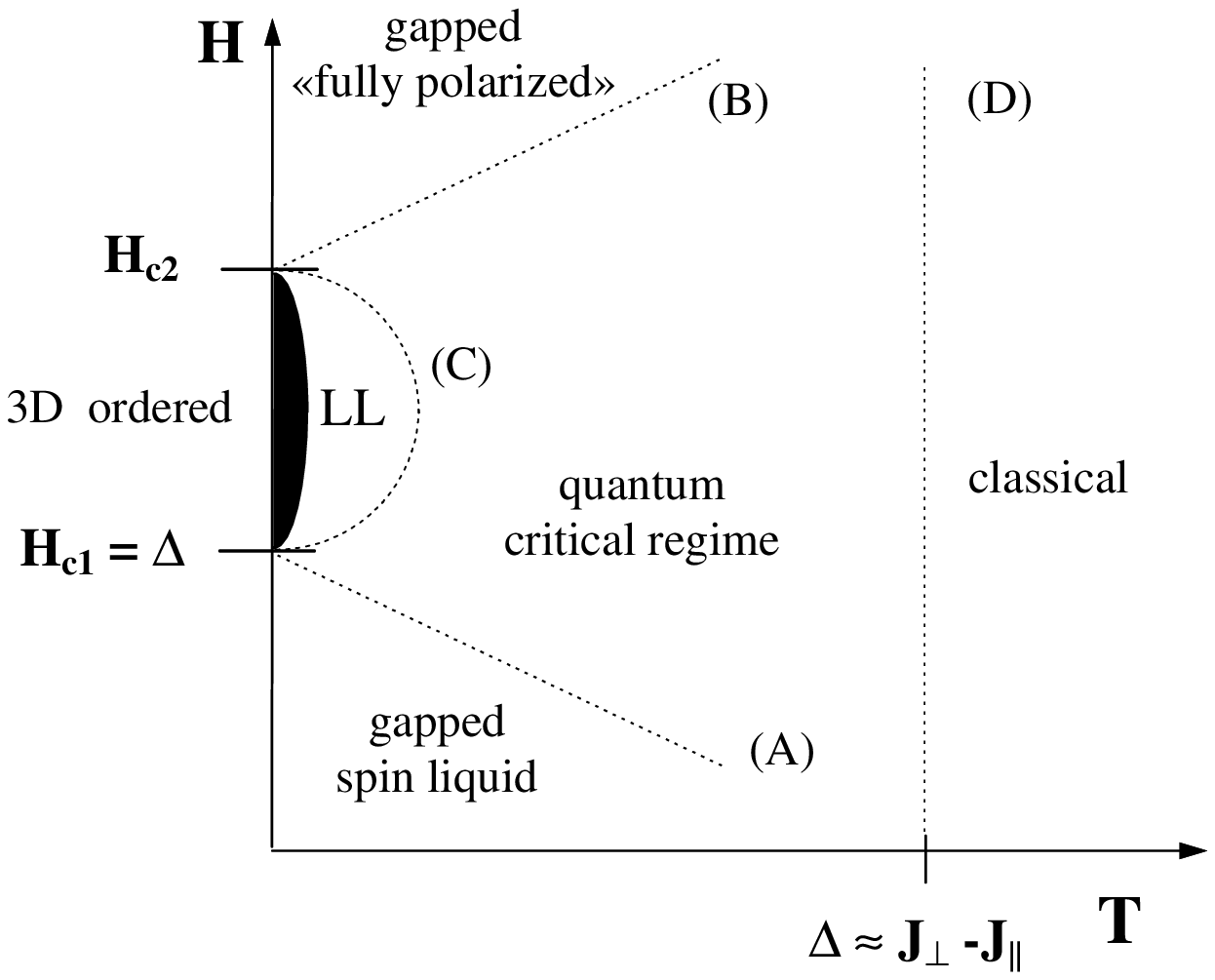}
\caption{Phase diagram of spin-ladders in a magnetic
field.  The magnetic field (vertical axis), can be thought as a tuning
parameter for the quantum coupling constant $g$.  Below $H_{c1}
\Leftrightarrow g_c$, the coupling constant $g$ exceeds $g_c$ ($H \le
H_{c1} \Leftrightarrow g \ge g_c$).  This phase has a singlet ground
state and an energy gap.  Above $H_{c1}$, the coupling constant $g$
drops below $g_c$:  the phase is magnetic and gapless and belongs to the
same universality class as the Heisenberg XXZ model.  The dotted lines
represent the crossover lines where $\xi_g \simeq \xi_T$ separating two
quantum regimes just described (where $\xi_g$ is the relevant quantum
correlation length) from the quantum critical regime dominated by
$\xi_T$.  For the system discussed in this paper the quantum critical
behavior at the upper critical field $H_{c2}$ is similar and will be
discussed in the next sections.  The shaded area between critical 
fields is a 3D ordered phase stabilized by transverse interactions (see 
Sec.~5).}
\label{fig1}
\end{figure}

To summarize, there are two-relevant lengthscales at a quantum critical
point, the quantum correlation length $\xi_g$ and the thermal length
$\xi_T$.  Depending on their relative values, different regimes exists.
They are represented graphically in Fig.~\ref{fig1}, on the H-T phase
diagram appropriate to spin-ladders.  The regions dominated by quantum 
effects are (a) the gapped spin-liquid phase below line A, (b) the XXZ 
or Luttinger liquid phase to the left of line C and (c) the gapped 
polarized phase above line B.  The quantum critical region is found 
to the right of these crossover line and extends down to $T=0$ at the 
critical fields $H_{c1}$ and $H_{c2}$.  Each regime will clearly be
identified using thermodynamic and $1/T_1$ NMR relaxation measurements
which allow to place precisely the crossover lines on this phase
diagram.

The differents sections are organized as follows:  a brief description
of several families of gapped antiferromagnets having an H-T phase
diagram similar to the one shown in Fig.~1 can be found in Sec.~2.  A
description of the structure and the interactions relevant to CuHpCl,
the 1D ladder system chosen for our case study, is given in Sec.~3.  A
mapping of strongly coupled ladder in a magnetic field onto the XXZ
Heisenberg model is introduced in Sec.~4.  It is used throughout the
rest of the paper to fit and interpret experimental data.  In Sec.~5,
the different phases shown in Fig.~1 are identified using high-field
magnetization data.  The XXZ model is used to fit the low temperature
data and to give a physical model for the ordered phase observed between
$H_{c1}$ and $H_{c2}$.  Section 6 is devoted to the dynamical processes
entering in the NMR relaxation.  The different regimes decribed in
Fig.~1 are presented in Sec.~7 through $T_1^{-1}$ measurements across 
the entire phase diagram.  The Luttinger liquid behavior between 
$H_{c1}$ and $H_{c2}$ is clearly seen for the first time and compared to
the XXZ models.  Finally the first scaling analysis
for a 1D quantum critical point is presented in Sec.~8 and the paper 
concludes with some new perspectives.
%
\section{Gapped 1-D antiferromagnets:  a broad universality class}
\begin{figure} 
\centering
\includegraphics[height=50mm, angle=0]{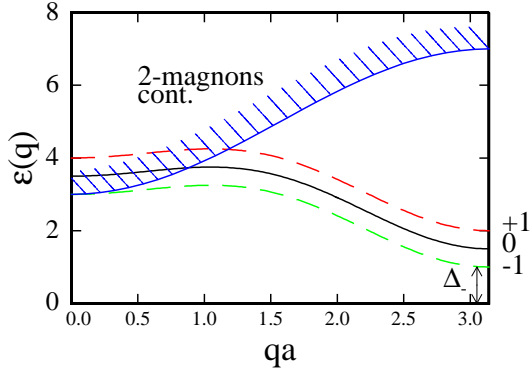}
\caption{The lowest lying excitations form a triplet branch, split by
the Zeeman field, or other anisotropic forces.  They are separated from
a unique singlet ground state by energy gaps $\Delta_+, \Delta_0,
\Delta_-$.  In absence of anisotropies, they are split by the Zeeman
energy, $\Delta_\pm = \Delta_0 \pm g\mu_B H$.  The shape of the
dispersion relation across the Brillouin zone depends on the exchange
interactions specific to each 1-D system.  In particular, the energy gap
may be at $qa=\pi$ (S=1 chains, ladders) or at $q=0$ (dimerized chains
with sufficiently large antiferromagnetic second-neighbor exchange). 
Similarly,
a two-''magnons'' continuum may intersect the triplet branch and have a
lower energy in some region of the Brillouin zone.  In this universality
class, there are no singlet states in the gap.}
\label{fig2} 
\end{figure}
\begin{figure} \centering
\includegraphics[height=50mm, angle=0]{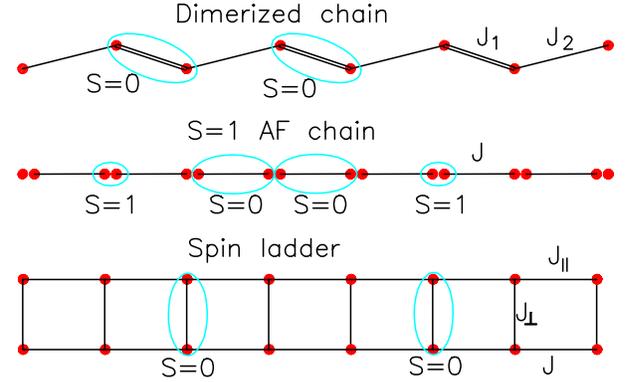}
\caption{Three families of gapped antiferromagnets.  Top, dimerized
chains:  the antiferromagnetic bond $J_1$ dominates $J_2$.  Singlet
valence bonds are preferentially localized on these bonds.  Middle,
integer spin chains.  The ground state can also be viewed as a product
of valence bonds\protect\cite{Affleck87}, by decomposing each spin 1 in
two spin-1/2 and forming valence bonds with the pair of spin-1/2 at the
extremity of each bonds (valence bond solid).  Bottom:  spin ladders.
In the strong coupling limit ($J_\perp \gg J_\parallel$), singlet bonds
are localized on the rungs of the ladder.}
\label{fig3}
\end{figure}

There are three known families of quasi 1-D materials belonging to the
same universality class, with an H-T phase diagram similar to Fig.~1.
They can all be described by the same quantum-field theory, the O(3)
non-linear $\sigma$-model\cite{Haldane83,Affleck89,Levy97} in a magnetic
field.  Their excitation spectrum\cite{Uhrig96} in a weak magnetic field
is represented in Fig.~\ref{fig2}.  The lower critical field $H_{c1}$,
is reached when the lowest energy gap $\Delta_-$ vanishes.  The upper
critical field $H_{c2}$ is usually reached when the highest energy state
of the $\epsilon_-(q)$ branch is below the singlet energy.  An
approximate representation of their singlet ground-states are
represented in Fig.~\ref{fig3}.

The know families of 1-D antiferromagnets in this universality class
are:\\
- (i) Quasi 1D-antiferromagnetic compounds with two alternating exchange
constants $J_1$ and $J_2$ have been known for over two decades to have a
gap (top of Fig.~\ref{fig3}).  For spin-1/2 alternating chains, the
spectrum is identical to Fig.~\ref{fig2} and in the strong coupling
limit ($J_1 \gg J_2$), the energy gap is $\Delta \approx J_1 -
\frac{J_2}{2} -\frac{3}{8}\frac{J_2^2}{J_1}$.  Very nice thermodynamic
studies\cite{Diederix79} of $Cu(NO_3)_2 . 2.5 H_2 0$ have identified the
existence of two critical fields $H_{c1} \approx J_1-J_2/2 \simeq 2.8$ T
and $H_{c2} \equiv 2 J_1 \simeq 4.5$ T. In spite of the modest value of
$H_{c1}$ and $H_{c2}$, dynamical properties close to these critical
points have never been thoroughly mapped out neither by NMR relaxation
measurements nor by neutron scattering.  Considering the interest in
quantum phase-transition, this interesting compound should be
revisited.\\
- (ii) Spin-1 Heisenberg antiferromagnetic chains\cite{Haldane83} with a
sufficiently weak planar anisotropy have in zero magnetic field a
triplet excitation branch separated by an energy gap $\Delta \approx
0.41 \times J$ from the unique singlet ground state.  The most
thoroughly studied compound in this family is NENP\cite{Renard87}.
Because of the presence of a planar anisotropy, this system has three
different lower critical fields (9.8, 13.3 and 14
Tesla)\cite{Ajiro89,Katsumata89} depending on the orientation of the
field with respect to the anisotropy axes.  The upper critical field
which has not been measured should exceed 86 Tesla.  Thermodynamic and
dynamical measurements\cite{Fujiwara93} have been carried out at the
lower critical field and provide very valuable insight on
zero-temperature phase transitions.\\
- (iii) Spin-ladders are quasi-1D structures where a finite number of
antiferromagnetic chains are coupled by a transverse antiferromagnetic
exchange.  Ladders with an odd-number of coupled chains are gapless and
belong to the same universality class as the spin-1/2 Heisenberg chain
\cite{Dagotto96}.  On the other hand, spin-1/2 ladders with an even
number of legs are gapped and form a singlet ground-state with short
ranged spin correlations (spin-liquid).  While several compounds with
ladder-like magnetic structure exist \cite{Azuma94,Hiroi95},
the only system where the quantum critical point $H_{c1}$ is
experimentally accessible is CuHpCl, a coordination compound made up by 
stacking binuclear molecules in a ladder structure\cite{Chiari90}.
Thermodynamic\cite{Chaboussant97a,Hammar98} and dynamic
quantities\cite{Chaboussant97b,Chaboussant98} have been measured over 
the entire phase diagram and give a relatively complete experimental 
picture of a zero-temperature phase transition.

We now describe its structure, and the relevant magnetic interactions 
in this material.
%
\section{CuHpCl, a 1-D spin ladder in the strong coupling limit}
\begin{figure}
\centering 
\includegraphics[height=40mm, angle=0]{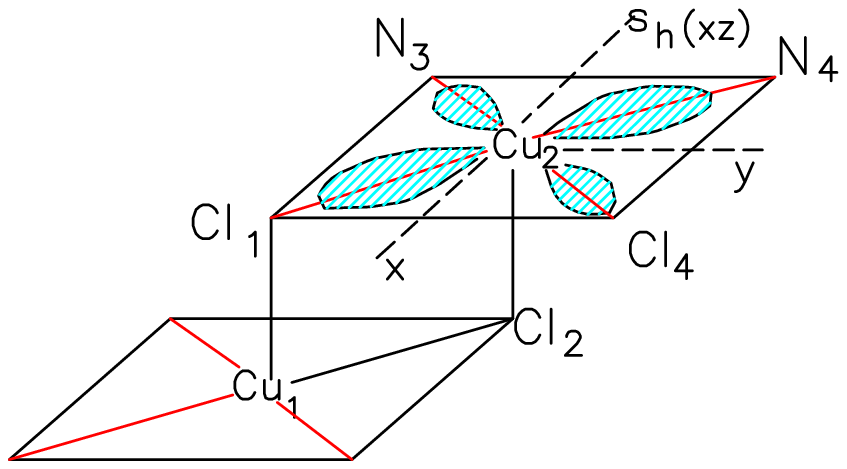}
\centering 
\includegraphics[height=85mm, angle=0]{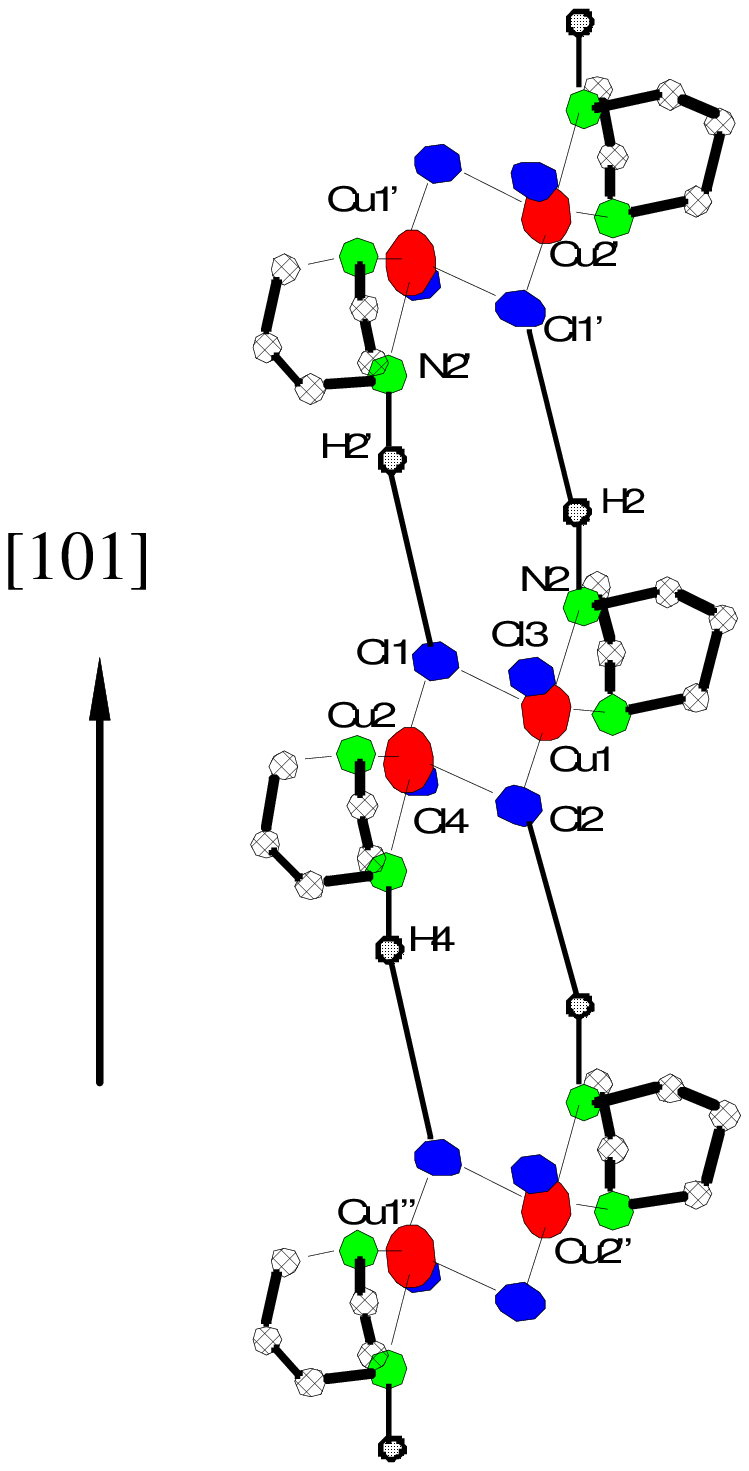}
\caption{Top: Representation of the parallel plane geometry
of the $Cu^{++}$ orbitals.  Bottom:  stacking of the molecular units in 
a ladder structure along the $[101]$ axis.}
\label{fig4}
\end{figure}
The molecular unit is a binuclear structure with two $Cu^{++}$ ions
(spin 1/2) each lying in a middle of two parallel distorted
square-structures\cite{Chiari90}.  At the vertices of each square, one 
finds two chlorine and two nitrogen ions as depicted in the top of Fig.
\ref{fig4}.  There are two super-exchange paths through the chlorine
ions $Cl_1$ and $Cl_2$ bridging the copper ions.  However, the $p_z$
orbitals of the chlorine ions are nearly orthogonal to the $Cu^{++}$
$d_{xy}$ orbital.  The resulting exchange constant between the
copper ions $J_\perp \approx 13.5\:K$, is found to be weaker than in
other materials with similar $Cu-Cu$ distance.  The organic rings, on
the outside of the ionic-core just described, contribute further to the
delocalization of the unpaired $Cu^{++}$ orbital.  Each molecular unit
stacks up in the [101] direction of this $P2_1/c$ molecular crystal (see
Fig.  \ref{fig4}).  In addition to the van-der-Vaals forces, the organic
rings allow a weak hydrogen bonding between molecules, along which a
super-exchange path can propagate.  In spite of the relatively large
intermolecular $Cu-Cu$ distance, the orbital overlaps are favorable and
lead to an intermolecular-exchange interaction $J_\parallel \approx
2.45\:K$ along the ladder.  Strictly speaking, due to the low symmetry,
there are small differences in $Cu-Cu$ distances along the ladder above
and below each molecular unit.  Because the bridging entities have the
same symmetries, these small differences are not expected to modulate
significantly the exchange ($J_\parallel$) along the legs.  The
translational invariance is nevertheless naturally broken, making
any additional lattice (spin-Peierls) instabilities less favorable. 
Comparison of
thermodynamic properties to numerical simulations\cite{Hawyard96} have
demonstrated that the other possible magnetic cross-bondings between the 
legs are weak and need not be considered.  On the other hand, it
is now clear that there is also a weak inter-ladder super-exchange,
probably also mediated by a weak hydrogen bonding between organic rings.
If it is not relevant in the gapped phases, it induces at low-T a
3D-ordered phase in the gapless region between $H_{c1}$ and 
$H_{c2}$\cite{Hammar98}.
While this phase has been observed in specific heat measurements, its
actual structure has not been determined experimentally.
%
\section{Hamiltonian representations of strongly-coupled spin-ladders}
For most purposes, it will be sufficient to consider a quasi-1D ladder
Hamiltonian in a magnetic field ${\mathcal H} = {\mathcal H}_{1} +
{\mathcal H}_{2} + {\mathcal H}_Z$, where
\begin{eqnarray} 
{\mathcal H}_{1} & = & J_\perp \sum_{i=1}^N {\vec S}_{2i-1} \cdot
{\vec S}_{2i} \\
{\mathcal H}_{2} & = & J_\parallel \sum_{i=1}^{2N} {\vec S}_i \cdot
{\vec S}_{i+2},
\label{eq2}
\end{eqnarray}
(even spins are on one leg and odd spins are on the other).  The weak
$g$-factor anisotropy ($g_\perp \approx 2.03, g_\parallel \approx 2.11$)
observed in EPR measurements\cite{Chaboussant97a} may be retained in the
Zeeman Hamiltonian, ${\mathcal H}_Z = \sum_{i=1,\alpha}^{2N} g_\alpha
\mu_B S_i^\alpha H_\alpha$.  In the strong coupling limit ($J_\perp \gg
J_\parallel$), it is possible to give a straightforward description of
the low-energy states in a magnetic field, treating ${\mathcal H}_{2}$
as a perturbation\cite{Totsuka98,Mila98}.  The eigenstates of ${\mathcal 
H}_{1}+{\mathcal H}_Z$
which describes isolated dimers in a magnetic field, are built from the
singlet $\sbond$ (valence bonding) and triplets $\tbond_{+1}$,
$\tbond_0$, $\tbond_{-1}$ (antibonding) on each rung.  Since we are
interested in the critical region, where the Zeeman energy is of the
order of $J_\perp$, it is legitimate to project ${\mathcal H}_{2}$ on
the restricted Hilbert space generated by the lowest dimers states,
$\sbond$ and $\tbond_{-1}$.  The matrix elements of ${\mathcal H}_{2}$
between neighboring dimers can be represented on this subspace by a $2
\times 2$ matrix, which is expressed in second-quantized notation as
\begin{equation}
{\mathcal H}_{2}^{\rm eff} = \frac{J_\parallel}{2} \sum_{r=1}^N 
\left( t^\dagger_r t_{r+1} + t_r t^\dagger_{r+1} + n_r n_{r+1} \right).
\label{H_eff}
\end{equation}
The fermionic operator $t^\dagger$ creates the triplet state
$\tbond_{-1}$ on bond $r \equiv [2i,2i+1]$ (only one triplet per bond is
allowed), while $t$ destroys a triplet, leaving a singlet on bond r. The
operator $n_r \equiv {t^\dagger}_r t_r$ counts the triplet occupation of
bond r. In the fermion language, the first two terms represent the
kinetic energy while the last term is a short range repulsion between
fermions.  Since $S^2$ and $S_z$ are good quantum numbers (${\vec S}$ is
the total spin), it is convenient to divide the Hilbert space into
sectors with a given value of $S_z$.  In the restricted Hilbert space,
each sector specifies the total fermionic occupation since
\begin{equation}
S_z = \sum_r n_r.
\end{equation}
The singlet sector is not coupled by ${\mathcal H}_{2}^{\rm eff}$ and
the singlet eigenstate remains the dimer product $|S=0\rangle=|\sbond
\sbond \ldots \rangle$ with energy $E_0=-\frac{3}{4}N J_\perp$.
This ground state energy may be compared to a serie 
expansion\footnote{When the full
Hilbert-space is retained, the strong coupling expansion for the singlet
ground state reads
\begin{eqnarray}
& &|S=0\rangle  = | \sbond \sbond \ldots \rangle +
\frac{J_\parallel}{J_\perp} \sum_r \\
& &|\sbond \ldots \sbond_{r-1} 
\frac{(\hbond +\xbond)_{r,r+1}}{\sqrt{2}} \sbond_{r+2} \ldots \rangle
+ {\mathcal O}\left( \frac{J_\parallel}{J_\perp} \right)^2, \nonumber \\
& &E_0 =-\frac{3}{4}N J_\perp \left[ 1 +
\frac{1}{2}\left(\frac{J_\parallel}{J_\perp} \right)^2 + \right]
\end{eqnarray}
where the same valence bond notation is used for all
singlets whether they lie along the legs or the rungs.}
in $J_\parallel/J_\perp$\cite{Reigrotzki94}.    For the parameters
appropriate to CuHpCl, the corresponding singlet energy
is only 1.6\% lower than in the previous estimate.  The reduction of
${\mathcal H_2}$ to Eq.~\ref{H_eff} is therefore appropriate to CuHpCl,
at least for qualitative answers.

In the $S=1$ (1-fermion) sector, the Hamiltonian ${\mathcal H}_{2}^{\rm
eff}$ can also be diagonalized in Fourier space, since the interaction
term $n_r n_{r+1}$ does not contribute.  The dispersion relation of this
triplet state is,
\begin{eqnarray}
E_{S_z=-1} (q) & = & J_\perp -g_z \mu_B H + J_\parallel \cos(qa), 
\label{E1q}\\
|1,q\rangle & = & \sum_r \exp(i q r)|\sbond \ldots \sbond_{r-1} \tbond_r
\sbond_{r+1} \ldots \rangle.
\end{eqnarray}
For fields below $g_z \mu_B H_{c1} \approx \Delta_0 = J_\perp -
J_\parallel$, there is an energy gap $\Delta_-=\Delta_0-g_z\mu_B H$
between the singlet and the $q=\pi/a$ minimum of the triplet branch, as
shown in Fig.  \ref{fig2}.  For fields above $H_{c1}$ or temperatures
above the gap $\Delta_-/\Boltz$, it is necessary to explore the energy
spectrum of ${\mathcal H}_2$ at finite fermion density.  In the
low-energy sector of the Hilbert space ($\sbond, \tbond_{-1}$), a finite
fermion density raises the energy of ${\mathcal H}_1+H_z$ with respect
to the ground state by an amount proportional to the triplet (fermion)
density
\begin{equation}
\delta E = (J_\perp -g_z \mu_B H) \sum n_r.
\end{equation}
In other words, $\mu \equiv J_\perp - g_z \mu_B H$ acts as the chemical
potential for the triplets.
\begin{figure}
\centering
\includegraphics[width=90mm, angle=90]{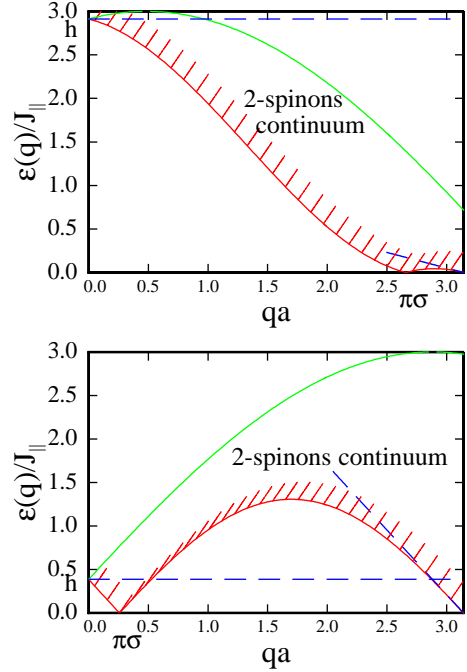}
\caption{Energy spectrum of transverse excitations in a magnetic field. 
Top panel: $H \ge H_{c1}$, the incommensurate wavevector is close to the
antiferromagnetic point.  Lower panel: $H \simeq (H_{c1}+H_{c2})/2$, the 
incommensurate wave-vector is close the zone center.  On both plots, 
the spin-stiffness $\rho_s \propto \sin(\pi-Q_\sigma)$ is 
the slope of the bottom edge of the spectrum at $q=\pi$.}
\label{fig5} 
\end{figure}

The low energy spectrum of ${\mathcal H}_2^{\rm eff}$ can be described
{\em exactly} on the restricted Hilbert space:  Eq.  (\ref{H_eff}) can
be recognized as the fermion representation of the $S=\frac{1}{2}$ XXZ
Heisenberg model
\begin{eqnarray}
{\mathcal H}_{\rm eff} &=& J_\parallel \sum_i \left( 
S_r^xS_{r+1}^x+S_r^yS_{r+1}^y
+\frac{1}{2}S_r^zS_{r+1}^z \right)\nonumber \\
& + & H_{\rm eff} \sum_r S_r^z +
N\frac{J_\perp}{8},
\end{eqnarray}
in an effective field
\begin{equation}
H_{\rm eff} = J_\perp + \frac{J_\parallel}{2} - g_z \mu_B H.
\end{equation}
$H_{\rm eff}$ is zero at the midpoint between $H_{c1}=J_\perp-J_\parallel$
and the upper critical field $H_{c2} \equiv J_\perp +
2J_\parallel$\cite{Chaboussant97a}.  The spin eigenstates of this
effective Hamiltonian are a representation of the triplet-singlet
subspace ($|\uparrow \rangle \equiv \tbond_{-1},|\downarrow \rangle
\equiv \sbond $) on each rung and have nothing to do with the original
spin-1/2.  In this model, the excitations which carry $\hbar/2$ angular
momentum (spinons), have a semionic character, i.e. can only be observed
in pairs.  In non-zero effective fields ($H_{\rm eff} \neq 0$) the
longitudinal and transverse excitations must be distinguished.
\begin{figure}
\centering
\includegraphics[height=55mm, angle=0]{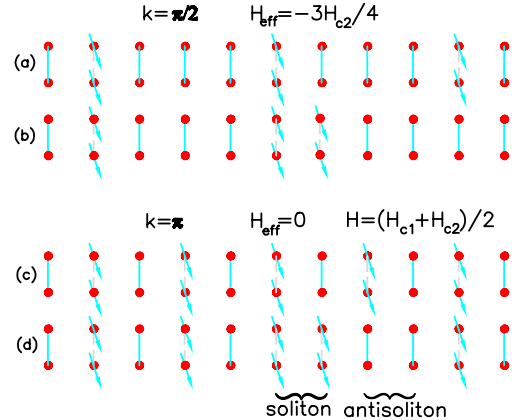}
\caption{The $k=\pi/2$ (a) and $k=\pi$ (c) ground states are
superposition of the singlet-triplet configurations shown.  The
wavevector $k$ refers to the wavevector in the XXZ representation.  In
ladder representation shown, this wavevector is $k/2$.  Excitations are
built as Bloch wave of the soliton-antisoliton pairs represented on (b)
and (d).}
\label{fig6} 
\end{figure}
The continuous spectrum of transverse excitations (1-magnon or
2-spinons) is represented on Fig.~\ref{fig5} at two different magnetic
fields.  At fields just above $H_{c1}$ (Fig.~\ref{fig5}, top panel) a
new minimum in the spectrum develops at $Q_\sigma a= \pi \sigma$, where
$\sigma=M(H_{\rm eff})/M_{sat}$ is the spin-polarization of the
XXZ-model in an effective field $H_{\rm eff}$.  The lower critical field
$H \approx H_{c1}$ correspond to the saturation field of the XXZ model,
$H_{\rm eff}^{sat}\approx 3J_\parallel/2$:
the spin-polarization $\sigma$ in $H_{\rm eff}^{\rm sat}$ is $-1$ and
the ladder magnetization $m$ is zero.  Just above $H_{c1}$, the soft
modes at $\pi/a$ and $Q_\sigma$ are very close:  the spin stiffness
$\rho_s \propto (1+\sigma) \propto m$ goes to zero and a large
low-energy spectral weight exists close to the antiferromagnetic point.
The situation is quite different close to $H_{\rm eff}=0$ (i.e.
half-way between $H_{c1}$ and $H_{c2}$), where the incommensurate
wave-vector $Q_\sigma$, is close to the zone center (Fig.~\ref{fig5},
lower panel).  For longitudinal fluctuations, the incommensurate minima
in the two cases considered are essentially interchanged with respect to
transverse fluctuations.

It is useful to ''translate'' the XXZ incommensurate states just
described into the valence bond representation of ladder states.
The pictorial images of the incommensurate ground states shown
in Fig.~\ref{fig6}, close to $J_\perp-J_\parallel/4$ and
$J_\perp+J_\parallel/2$, are appropriate on short lengthscales (quantum
fluctuations destroy the periodicity $Q_\sigma$ on long lengthscales).
Low energy excitation above the ground state are Bloch waves of the
soliton-antisoliton defects depicted on Fig.~\ref{fig6}:  this builds a
coherent superposition $\alpha \sbond + \beta \tbond_{-1}$ (transverse
fluctuation) at wavevector $Q_\sigma$.

Since so many exact results are known for the XXZ model, the mapping 
discussed here will prove to be extremely useful in the analysis 
experimental data.  The high field magnetization of CuHpCl 
clearly establishes the correspondence with the phase diagram show in 
Fig.~\ref{fig1}.
%
\section{Identification of the different phases with  high-field
magnetization measurements}
The weak g-factor anisotropy and the monoclinic symmetry of this
crystal, allow a straightforward determination of the magnetization by
torque magnetometry:  if the field is applied along the ${\hat z}$ axis,
which does not coincide with the principal axes ${\hat a}, {\hat b},
{\hat c}$, the magnetization is not collinear with H, and a torque $\tau
\propto M \times H$ can be measured.  The magnetization curves of a $100
\mu g$ monocrystal have been measured with an ultrasensitive AC torque
magnetometer\cite{Crowell96,Chaboussant97a}.
\begin{figure}
\centering
\includegraphics[width=78mm, angle=90]{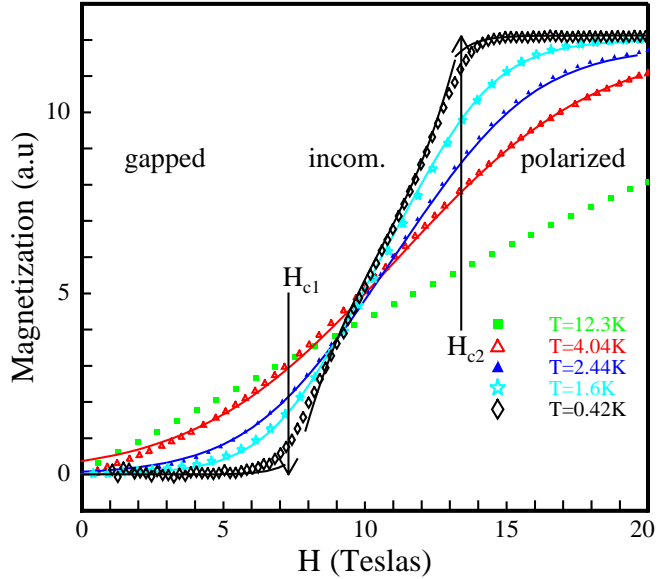}
\caption{Magnetization of CuHpCl between 0 and 20 T, at different 
temperatures.  The symbols are experimental data points, while the 
solid lines are the fits to the XXZ model described in the text, using
the procedure outlined in appendix B. From the lowest temperature
curve, two critical fields $H_{c1}$ and $H_{c2}$ are identified.  Below
$H_{c1}$, the system is unmagnetizable, i.e.  forms a singlet ground
state separated from $S \neq 0$ excitations.  Above $H_{c2}$, the system
is fully polarized.  The ''incommensurate phase'' is between $H_{c1}$
and $H_{c2}$.}
\label{fig7}
\end{figure}
It is straightforward to identify the critical field $H_{c1}$ and
$H_{c2}$ from the $T=0.42 K$ magnetization curve shown in
Fig.~\ref{fig7}.  Below $H_{c1}$, the magnetization is thermally
activated with an effective gap $\Delta_-$ (this is shown in
Fig.~\ref{fig8}).  Similarly, above $H_{c2}$, the deviation from the
saturated magnetization are activated with an energy gap $g_z\mu_B
(H-H_{c2})$ (see NMR section).  In appendix A, the energy spectrum of
excitations carrying one unit less angular momentum than the fully
polarized state are determined.  In a ladder, there are two spins in the
unit cell, and hence two spin-wave modes.  Since the polarized phase is
unstable when the lowest energy of these spin-waves drops below the
energy of the polarized state, the upper critical field can be specified
exactly
\begin{equation}
g_z \mu_B H_{c2} = J_\perp + 2 J_\parallel. \label{Hc2}
\end{equation}
The observed behavior of the magnetization in the different field
regions coincides precisely with the zero temperature phases specified
on the y-axis of Fig.  1. Quantitatively, the exchange parameters
$J_\perp \approx 13.5\:K$, $J_\parallel \approx 2.45\:K$ are most
accurately determined from the values of $H_{c1}$ and $H_{c2}$.
$J_\perp$ can be identified independently as the magnetic field at which
the NMR relaxation rate is maximum in the high temperature limit (see
Sec.  7).  These numbers have also been checked against (a) the low and
high temperature dependence of the
susceptibility\cite{Chaboussant97a,Weihong97}, (b) the gap suppression of
the low temperature specific heat and (c) the overall bandwidth ($2
J_\parallel$) of the triplet branch (H=0)\cite{Hammar98} measured by
neutron scattering.  In a numerical study of the ladder magnetization,
the presence of a weak cross-exchange coupling between legs has also
been investigated\cite{Hawyard96}.  The conclusion is that this coupling
is weak and if non-zero, ferromagnetic.  Considering all the
experimental and numerical uncertainties, it seems at present 
unnecessary to keep any additional exchange coupling in the analysis.

The thermodynamic properties for the XXZ model can be computed exactly
by the Bethe Ansatz\cite{Yang66,Takahashi72} and the magnetization
curves have been evaluated numerically using the procedure described in
Appendix B. At temperature below $J_\perp$,\cite{hightemp} the results
(Fig.~\ref{fig7}, solid lines) agree very well with the experimental
magnetization.  Considering that there are no adjustable parameters, the
mapping of strongly-coupled spin-ladders onto the XXZ model appears to
be excellent.  In particular, the quantum critical behavior of ladders 
at $H_{c1}$ is acurately reproduced by this model, a point which is 
emphasized further in Sec. 8 by constructing explicitly the
scaling plots for the magnetization. In the gapped phase, 
the XXZ mapping
progressively loses its validity at small Zeeman splitting (compared to
$\Delta$)\cite{hightemp}.  In this limit, it is instructive to compare
the temperature dependent magnetization to the free-fer\-mion model
proposed by Troyer et al.\cite{Troyer94,Chaboussant97a} where
\begin{equation}
\frac{M}{M_{sat}}=\frac{2z(\beta)\sinh(\beta g_z\mu_B 
H)}{1+[2\cosh\beta g_z \mu_B H]z(\beta)}
\end{equation}
and $z(\beta) \approx \exp(-\beta J_\perp) I_0(\beta J_\parallel)$ In
this model, the statistical weights are adjusted in order to reproduce
the full spin entropy at high temperature, but the nearest-neighbor
repulsion is ignored.  Fig~\ref{fig8} shows that, at high temperature, 
the experimental magnetization is systematically higher that inferred
by the free-fermion model (solid lines).  At finite temperature,
interactions between fermions increase the value of the chemical
potential (which is negative in the gapped phase, $\mu \ge
-|\Delta_-|$).  Hence at higher temperature, the effective gap ($|\mu|$)
becomes smaller and a higher fermion density is possible.  The role of
interactions will be further emphasized with the identification of the
NMR relaxation processes:  the dominant relaxation channel (staggered
process) when $H \rightarrow H_{c1}$ would not be possible without
interactions.
\begin{figure}
\centering
\includegraphics[width=60mm, angle=90]{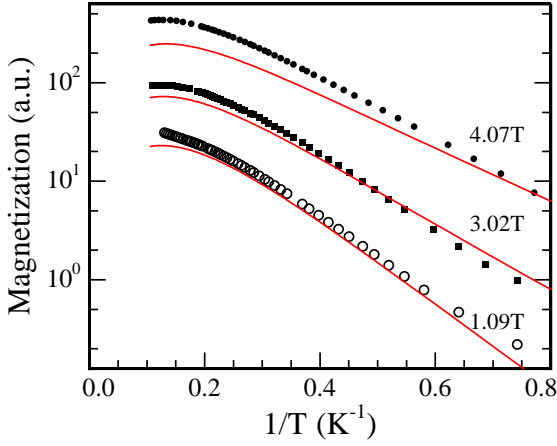}
\caption{Magnetization of CuHpCl in the gapped phase as a function of
$1/T$, at different magnetic fields.  The experimental points are
compared to the free fermion model of Troyer (solid lines).  A
systematic deviation from the free-fermion model is observable when
their density becomes significant:  this emphasize the role of the short
range repulsion between fermions.}
\label{fig8}
\end{figure}
%
%
\subsection{Ordered phase}
Down to $T=0.1\:K$, magnetization curves show no plateaux nor slope
changes which could indicate a 3D ordering transition.  The 1D models
appear to give a precise account of the magnetization at all
temperature.  On the other hand, small but sharp peaks in the specific
heat have been observed above $H_{c1}$ at low
temperature\cite{Hammar98,Calemzuck98}.  After integration of the
specific heat at constant field, it is found that the entropy per spin
associated to this transition is very small ($<10^{-2} \Boltz$/spin).
In light of these experimental facts, a {\em second order} phase
transition appears at low-T between $H_{c1}$ and $H_{c2}$, involving a
small change in spin-entropy and no detectable change in magnetization.
\begin{figure}
\centering
\includegraphics[height=50mm]{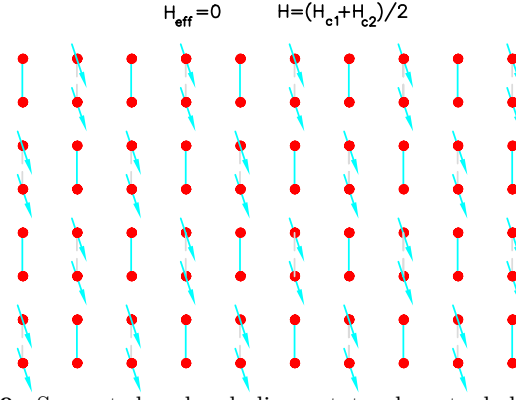}
\caption{Suggested ordered dimer-state close to half-filling:  triplet
and singlet bonds form a 3D antiferromagnetic lattice.}
\label{fig9}
\end{figure}

When 3D coupling are ignored, the XXZ mapping (Sec.  4) gives a
representation in terms of an interacting 1D fluid of spinless fermions 
(Luttinger liquid).  Since there is no magnetization change, the 3D
transition takes place at constant fermion density.  A very common 3D
instability for a 1D Luttinger liquid is a charge density wave ordering.
In a valence-bond language, this transition can be viewed as a
valence-bond ordering of the 1D $\tbond_{-1}$ states in a 3D-lattice.
This would hardly affect the magnetization which measures the density of
$\tbond_{-1}$ bonds (fermion density) while the quenching of their 
kinetic energy would be manifest in the specific heat.  For a 3D-charge
ordering, repulsive interactions between fermions are usually necessary.
Antiferromagnetic superexchange between ladders, introduce very naturally
an additional repulsion between fermions on different ladders.
This antiferromagnetic super-exchange may be represented by,
\begin{equation} 
{\mathcal H}_{inter}=J_{inter} \sum_{\langle k,l \rangle} {\vec S}_k
\cdot {\vec S}_l
\label{Hinter}
\end{equation} 
where the sum $\langle k,l \rangle$ is carried over nearest-neighbor
spins belonging to different ladders.  The physics of a low-T transition
should be described by the projection of (\ref{Hinter}) on the
restricted Hilbert space, i.e. 
\begin{equation}
{\mathcal H}_{inter}^{\rm eff} = -\frac{J_{inter}}{4} \sum_{\langle k,l
\rangle} \left( t^\dagger_k t_l + t_k t^\dagger_l - n_k n_l \right),
\end{equation} 
which has, up to a sign, the same form as ${\mathcal H}_2^{\rm
eff}$.  A gauge transformation, switching the phase of hopping 
operators $t_{2r+1} \rightarrow -t_{2r+1}$ every other rungs, restores 
an effective antiferromagnetic coupling in the spin model.  In the 
spinless fermion model, a transition to a
charge density wave state can only be established close to half filling
[$H=(H_{c1}+H_{c2})/2$]\cite{Georges98}.  It is not known whether this
transition persist at low fermion densities.  Since this model arise in
many context, it will be important to determine its complete phase
diagram.  Of particular relevance is the commensurate or incommensurate
nature of the 3D charge density.  The physics of this model is in fact
relevant to almost all quasi-1D quantum magnets.  For example, in the
spin-Peierls compounds $CuGeO_3$, there is also a transition to a 3D
incomensurate phase, with no change in total magnetization.  At the
transition, only the local distribution of magnetization
changes\cite{Fagot96}.  E\ss ler and Tsvelik\cite{Essler97} have shown
that the 3D phonon-couplings present in this family of materials can be
represented as a transverse exchange between chains in an effective
magnetic Hamiltonian.  From the point of view of magnetism, this
''charge density wave'' ordering cannot be distinguished from a real
spin-Peierls transition recently proposed by several
authors\cite{Calemzuck98}.  It is therefore natural to expect ordered
phases with similar structures in all compounds within this universality
class.

From the point of view of magnetism, this is an original magnetic state,
with a 3D ordered structure of valence bonds.  Fig.~\ref{fig9} gives a 
pictorial representation of this state at half-filling 
($(H=H_{c1}+H_{c2})/2$).  From this discussion, it is clear that further
experimental and theoretical studies of the 3D ordering of strongly 
coupled ladder are called for.
%
\section{Assigment of NMR lines and identification of the dynamical 
relaxation processes in the gapped phase}
\begin{figure}
\centering
\includegraphics[height=60mm]{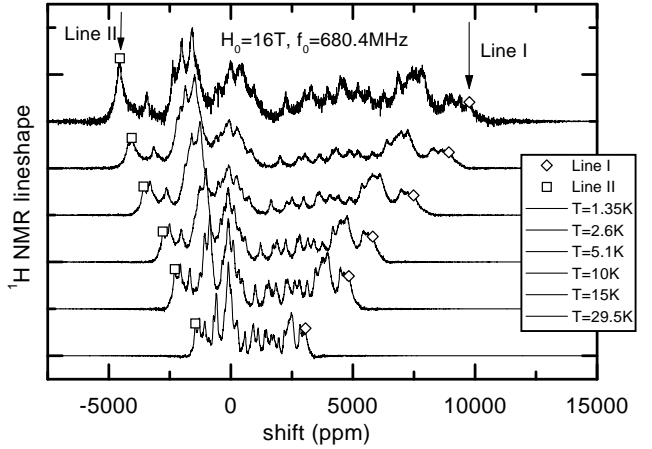}
\caption{Temperature dependence of the NMR spectrum measured at 16T.
The intensities have been normalized to compensate for the gap
suppression.  Most $T_1^{-1}$ measurements have been carried out on Line
I and II, which are always located at the extremities of the spectrum
for all fields and temperatures.  They can be identified with nuclei
$H_{14}\:\&\:  H_{23}$ (line I) and nuclei $H_2$ (line II).}
\label{fig10}
\end{figure}

NMR is an ideal tool to probe the low-energy dynamics of quantum
magnets\cite{Chaboussant97b}.  When nuclei (here protons) are located at
different sites than the electronic spins, the interaction between
electronic and nuclear spins are mostly dipolar:  the dipolar field
${\vec h}_{ij}(t)$ produced by the electronic spin $i$ on the nuclear
spin $j$ serves a probe for the dynamical properties.  The time-averaged
$z$-component $\sum_i \langle h^z_{ij} \rangle$ of this local field
shifts the value of the magnetic field felt by the nucleus by an amount
proportional to the local electronic susceptibility $\chi_i$
\begin{equation}
K_j = \sum_{i} A_{ij} \chi_i.
\end{equation}
$K_j$ depends on the position of the nuclei in the unit-cell through the
dipolar sum
\begin{equation}
A_{ij} \propto -\gamma_e \gamma_n \hbar^2 \frac{1-3\cos^2
\theta_{ij}}{|r_{ij}|^3}
\label{Aij}
\end{equation}
where $\theta_{ij}$ is the angle between ${\vec r}_{ij}$ and $\vec H$.
$K_j$ will be positive if ${\vec h}_j$ is mostly parallel to H, and
negative if its antiparallel.  $CuHpCl$ contains 24 protons spins in the
unit cell:  in the NMR spectrum at $16\:T$ (i.e.  in the polarized
phase) shown in Fig.~\ref{fig10}, there are indeed 24 resolved lines.
Their position depends on temperature and follow the T-dependence of the
local magnetization.  For 16 nuclei the hyperfine shift $K_j$ is
positive and negative for the 8 remaining.  For proton sites which are
nearly equivalents, the lines are close together as the local fields
(and their fluctuations) are nearly identical.  
Using the proton positions calculated from X-ray data, it is possible to
determine the dipolar
sum on each site and assign it to a corresponding line.  For example,
the outermost line labeled I arise from nuclei $H_{14}\:\&\:  H_{23}$
($A_I \approx 2950\:G$) while line II arise from proton $H_2$ ($A_{II}
\approx -2400\:G$).

\begin{figure} 
\centering
\includegraphics[width=70mm,angle=90]{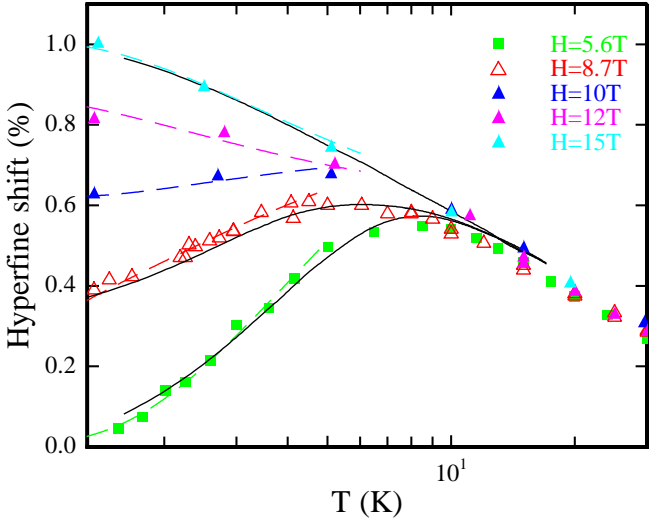}
\caption{Temperature dependence of the hyperfine shift at different 
magnetic fields.  The thermodynamic measurement of $M/H$
at H=5, 8 and 15T are
plotted on the same graph as solid lines.  The comparison of the NMR 
shift to the value of $M/H$ inferred from the XXZ mapping (Sec. 4) at 
low temperature are drawn as dotted lines.  At 8.7 T, in the gapless
XXZ-phase, there is first an increase of the local magnetization with
temperature, an unusual behavior for a magnetic state.}
\label{fig11}
\end{figure}

The hyperfine shift is also a measure of the local electronic spin
susceptibility $M_j/H$.  Its temperature dependence is shown in
Fig.~\ref{fig11}.  Below $H_{c1}$ there is, at low temperature, an
exponential drop of the hyperfine shift consistent with an activated
behavior with a characteristic energy gap $\Delta_-$.  The dependence
observed at $8.7\:T$, just above $H_{c1}$, shows an increase of the
local magnetization as the temperature is raised, a very unusual
behavior for a magnetic phase (e.g.  the magnetization of canted
antiferromagnets always decrease with temperature).  At low temperature
($\Boltz T < J_\perp$) it is meaningful to use the XXZ representation,
where the system can be viewed as a Luttinger liquid of triplets:  in
this limit, the presence of a continuum of longitudinal excitation
carrying an angular momentum $\hbar$ (see Sec.~4) contributes to an
increase of the triplet occupation with temperature.  At higher fields,
the density of states at small wavevectors gets smaller and a more
classical behavior is recovered.  In quantitative term, the hyperfine
shift observed below $6K \approx J_\perp/2$ agrees well with the XXZ
mapping and the thermodynamic measurements of $M/H$ (see
Fig.~\ref{fig11}).

Temporal fluctuations of the local fields ${\vec h}_{ij}(t)$ at the
nuclear precession frequency (essentially zero energy) provide the
dominant relaxation channel for nuclear spins.  The longitudinal
spin-lattice relaxation rate of the nucleus i
\begin{eqnarray}
\frac{1}{T_1} \bigg\vert_i & = & \int \exp(-i \omega_n t) dt \sum_j 
\langle h_{ij}^+ (t) h_{ij}^-(0) \rangle \\
& = & \frac{(\gamma_n\gamma_e \hbar)^2}{2} \sum_{q} \left[ F_\perp^i 
(q) {\mathcal S}_\perp (q, \omega_n) \right. \nonumber \\
& + & \left. F_z^i (q) {\mathcal S}_z(q, \omega_n) \right]
\end{eqnarray}
is sensitive to the transverse {\em and} longitudinal structure factors
\begin{eqnarray}
{\mathcal S}_\perp (q, \omega_n) & = & \int \exp(-i\omega_n t)dt \langle 
S_q^+ (t) S_{-q}^- (0) \rangle, \\ 
{\mathcal S}_z (q, \omega_n) & = & \int \exp(-i\omega_n t)dt \langle 
S_q^z (t) S_{-q}^z (0) \rangle
\end{eqnarray}
through the form factors $F_z^i$ and $F_\perp^i$.  These quantities are
simply the Fourier transform of $|A_{ij}|^2$ defined in Eq.~\ref{Aij}.
They are most easily computed in real space as geometrical dipolar sums:
the longitudinal and transverse components of the resultant local field
on site i, ${\vec h}_i = \sum_j {\vec h}_{ij}$, depend on the actual
position of the nucleus i in the unit cell.  Depending on the proton
site selected, the relative magnitude of the form factors $F_z^i$ and
$F_\perp^i$ can change by one order of magnitude:  this provides a
unique way to measure separately {\em all} components of the structure
factor at $\omega=\omega_n$.
\begin{figure}
\centering
\includegraphics[height=44mm]{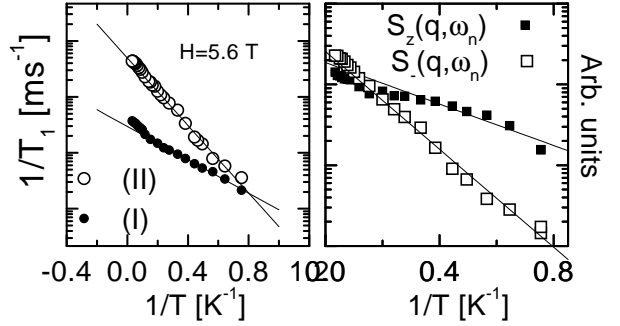}
\caption{Temperature dependence of the $T_1^{-1}$ relaxation rate of
sites I and II in the gapped phase ($H=5.6\:T$).  At high temperature
the relaxation at site II is an order of magnitude faster than at site
I, but decreases exponentially at lower temperature with an activation
energy $\Delta_{II} \approx 6.8\:K$, twice as large as for site I,
$\Delta_{I} \approx 3.4\:K$.}
\label{fig12}
\end{figure}

Fig.\ref{fig12} illustrates how different the relaxation on different 
proton sites can be in the gapped phase ($H=5.6\:T$).   If the 
$T_1^{-1}$ behavior of both lines I and II are activated,  
the activation energy for line II, $\Delta_{II} \approx 6.8\:K$, is 
twice as large as the activation energy for line I, $\Delta_{I} \approx 
3.4\:K$.  In a field of $5.6\:T$, the smallest activation energy is 
$\Delta_- = \Delta -g\mu_B H \ \approx 3.0\: K$ between the triplet 
branch $|1,q\rangle$ and the ground state, close to the measured 
energy $\Delta_{I}$ for line I.

The contribution of one-magnon states $|1,q\rangle$ (cf.  Eq.~\ref{E1q})
to the structure factor are proportional to $\delta (\omega-E_1(q))$:
their spectral weight at the nuclear frequency is zero.  On the other
hand, these states are by no-means exact eigenstates and a number of
low-energy scattering processes between magnons can generate a finite
spectral weight at low energy.  Among them, there are (i) finite matrix
elements of ${\mathcal H}_2$ between triplet states which are ignored in
the reduction to the effective Hamiltonian ${\mathcal H}_2$, (ii)
density-density interactions (last term in Eq.~\ref{H_eff}), which
contribute as the square of the density of thermally excited magnons,
(iii) other interactions such as interchain couplings and impurity
scattering.
\begin{figure}
\centering
\includegraphics[width=45mm, angle=90]{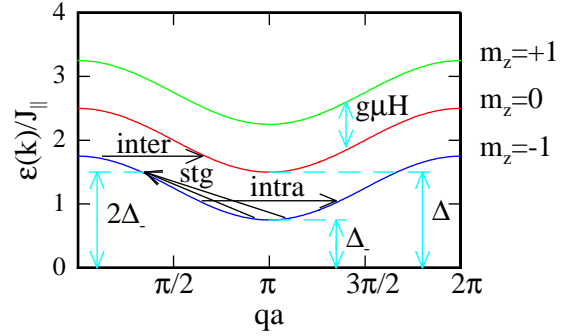}
\caption{Schematic representation of multi-magnon scattering processes 
which can drive the NMR relaxation.  The intrabranch processes leave 
$m_z$ unchanged and contribute only to ${\mathcal S}_z$.  The
interbranch process which involve a spin flip, contribute to ${\mathcal 
S}_\perp$ and are quenched one the Zeeman energy ($7.6\:K$ at 
$5.6\:T$) exceed the triplet bandwidth ($2J_\parallel \approx 5\:K$).
In the staggered process, two magnons from the bottom of the band 
$-1$ are scattered into a magnon state at twice the energy: it is 
governed by the square of the magnon density $n_-^2 \propto 
\exp(2\Delta_-)$.}
\label{fig13}
\end{figure}

It is instructive to follow the classification of low-energy processes 
proposed by Sagi and Affleck\cite{Sagi96} in the context of Haldane 
$S=1$ spin-chains.  Only three relevant channels need to be 
examined\\
- (i) The simplest processes are the spin-conserving two-magnons
processes (intrabranch) represented in Fig.~\ref{fig13}:  the magnons
states $\pi-q/2$ and $\pi+q/2$ within a magnon branch of a given $m_z$
are coupled by the hyperfine interaction which relaxes the nuclear
spins.  At low temperature, only states in the $m_z=-1$ branch are
thermally occupied with a density $\propto \exp(-\Delta_-/\Boltz T)$.
This occupation factor dominates the temperature dependence of this
relaxation channel.  In this limit, this process contributes
to the longitudinal structure factor ${\mathcal S}(\omega_n\approx 0, q
\approx 0)$ (no spin-flip).\\
- (ii) The spin non-conserving processes (interbranch) couple magnons
with energies $>\Delta$ in the $m_z=0$ and $m_z=-1$ branches.  They
contribute to the transverse structure factor (there is a spin-flip),
with an activation energy set by the full gap $\Delta$.  When the Zeeman
energy exceeds the one-magnon bandwidth ($2J_\parallel$), which is the
case at $5.6\:T$, this process is completely quenched since there are no
states left in $m_z=0$ and the $m_z=-1$ branches with the same
energies.\\
- (iii) Since the magnons states used do not constitute a real
representation for the eigenstates of the full Hamiltonian, various
processes quadratic in the magnon-density can occur.  The lowest order
process contributing to the transverse structure factor is a
three-magnon process where two-occupied magnons at the bottom of the
$m_z=-1$ band are scattered into a magnon with twice the energy via a
large momentum transfer.  The extra angular momentum being absorbed by
the nuclear spin, this process contributes to the transverse structure
factor ${\mathcal S}_\perp (\omega_n \approx 0, q )$, where $q$ is large
and will be taken as $\pi$ in the rest of the analysis.  This process is
governed by a quadratic thermal occupation factor $\propto n^2 =
\exp(-2\Delta_-/\Boltz T)$ and requires to have a final state in the
$m_z=-1$ branch available at energy $2\Delta_-$:  this is the case when
the bandwidth $2 J_\parallel$ exceeds the gap $\Delta-$, i.e.
sufficiently close to $H_{c1}$.  There are other relevant quadratic
processes:  four magnons scattering (Eq.~\ref{H_eff}) processes at the
bottom of the $m_z=-1$ band have the same temperature dependence but are
spin-conserving and hence enter only in the longitudinal structure
factor.

To summarize, the dominant processes in an intermediate field range 
are\\
- for the longitudinal structure factor ${\mathcal S}_z$, the
intrabranch two-magnon process, ${\mathcal S}_z \propto n_- \propto
\exp(-\Delta_-/\Boltz T)$\\
-for the transverse structure factor ${\mathcal S}_\perp$, the 3-magnons
staggered process represented in Fig.~\ref{fig13}, ${\mathcal S}_\perp 
\propto n_-^2 \propto \exp(-2\Delta_-/\Boltz T)$.

In this simple picture, two numbers ${\mathcal S}_z(q=0)$ and 
${\mathcal S}_\perp(q \approx \pi)$ are sufficient to specify the
$T_1^{-1}$ relaxation of all lines at a given field and temperature
\begin{equation}
\frac{1}{T_1}\bigg\vert_i \propto F^i_z(0){\mathcal S}_z+F^i_\perp (\pi) 
{\mathcal S}_\perp.
\label{linear}
\end{equation}
The value of the form factors $F$ appropriate for line I and II are
given in Table 1. Since $F_\perp^{II} \gg F_z^{II}$, line II is
dominated by the transverse structure factor, i.e.  is governed by the
staggered processes $\propto \exp (2\Delta_-/\Boltz T)$ in this field
range:  this is fully consistent the observed activation energy.  While
longitudinal and transverse fluctuations contribute to line I, the
linear system (\ref{linear}) can be solved explicitly and the different
component of the structure factor, plotted in Fig.~\ref{fig12}, show
very clearly the two different activation energies, corresponding to the
two dominant processes.

\begin{table}
\centering
\caption{The form factors for lines I and II, computed as dipolar sums,
are expressed in units of $10^{-4}$\AA$^{-6}$.  The uncertainties 
which are of the order of 20\%-30\% have a number of origins and
are discussed elswhere\protect\cite{Chaboussant97b}.}
\begin{tabular}
{|l|c|c|} \hline & $F_z(0)$ & $F_\perp(\pi)$ \\ \hline \hline
Line I & 13 & 6 \\ \hline
Line II &  4 & 70 \\ \hline
\end{tabular}
\end{table}

At high temperature, when the thermally excited fermion density is 
important, the staggered processes dominate by an order 
of magnitude.  Since the fermion density is likewise large above the 
critical field $H_{c1}$, staggered processes dominate the relaxation in 
the XXZ phase, a result which is consistent with all theoretical 
analysis\cite{Sagi96,Chitra97,Sachdev94} and the data presented in 
Fig.~\ref{fig15}.

\begin{figure}
\centering
\includegraphics[height=40mm, angle=0]{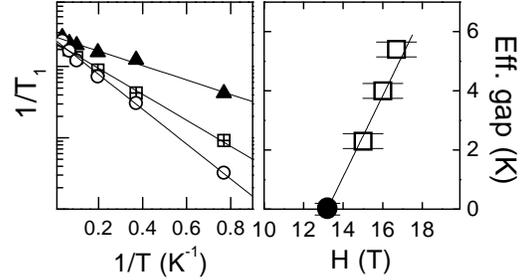}
\caption{Left: activated behavior of $T^{-1}$ (line I) in the high field 
phase.  Right: effective gap as a function of magnetic field.  The 
intercept on the field axis coincide with the value of 
$H_{c2}=13.2\:T$.}
\label{fig14}
\end{figure}

The relaxation rate  $T^{-1}$ in the polarized phase (above $H_{c2}$) 
is also activated as shown in Fig.\ref{fig14}.  The measured activation
energy is close to $g\mu_B(H-H_{c2})$ in good agreement with the
exact low energy spectrum described in appendix A. Hence, this polarized
phase has no Goldstone mode and in this sense is not a ferromagnet.
There are two magnons modes above the polarized state (two spins per
unit cell):  hence, the structure of the spectrum above $H_{c2}$ and
below $H_{c1}$ is qualitatively different, indicating that the XXZ model
which reproduces correctly the low-energy behavior in the gapless phase,
cannot be taken literally over the entire phase diagram.

%
\section{Incommensurate phase and quantum critical regime}
\begin{figure}
\centering
\includegraphics[height=95mm, angle=0]{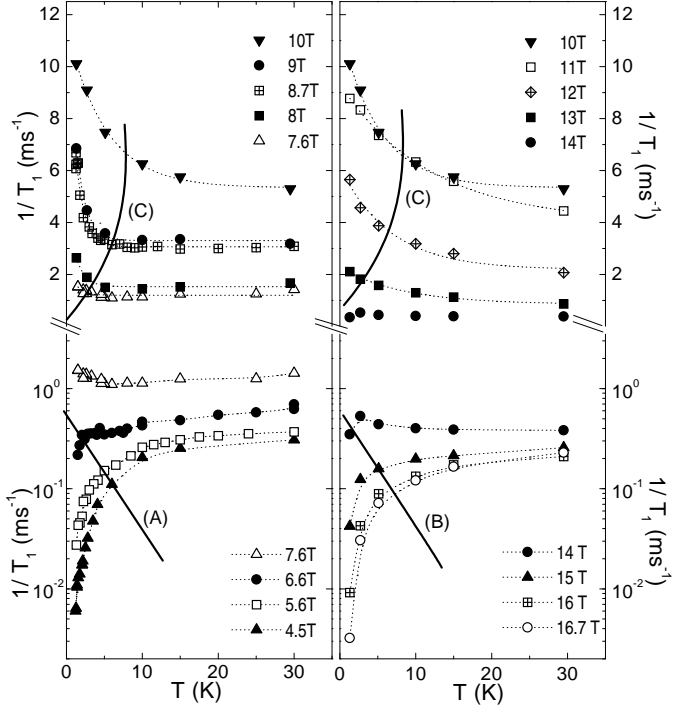}
\caption{Left panel :  temperature dependence of $1/T_{1}$ through the
critical field $H_{c1}$.  In the lower part, the $T_1^{-1}$ dependence
in the singlet gapped phase ($H < H_{c1}$) is displayed, while its
behavior in the magnetic phase ($H>H_{c1}$) is shown in the upper part.
The dashed lines are guides to the eyes.  Lines labeled (A) and (C)
correspond to the crossover lines of Fig.~\protect\ref{fig1}.  Right
panel:  temperature dependence of $1/T_{1}$ through the critical field
$H_{c2}$.  The upper part shows the behavior in the magnetic phase below
$H_{c2}$ while the lower part is in the "fully polarized" gapped phase
($H > H_{c2}$).}
\label{fig15}
\end{figure}
Fig.~\ref{fig15} gives the overall temperature and field dependence of
the NMR relaxation rate (line I) through the entire phase diagram.  In
the left panel, the dependence of the $T_1^{-1}$ rate through the lower
critical field is displayed.  In the lower part, the behavior in the
gapped phase is reproduced over a broader temperature range.  Two
distinct regimes can immediately be recognized.  When the temperature is
raised at constant field through the effective gap $\Delta_-$, the
exponential suppression of the $T_1^{-1}$ crosses over to the
temperature independent value of the relaxation rate expected in a
classical system.  Comparing with Fig.~\ref{fig1}, this crossover can be
recognized as line (A) (defined as $\Boltz T = \Delta_-$) separating the
gapped phase, controlled by quantum fluctuation, and the quantum
critical phase where $\xi_T$ is the only relevant lengthscale.  When the
field is raised above $H_{c1}$, the behavior at high temperature is
qualitatively the same but at low temperature $T_1^{-1}$ starts
to diverge.  Again, it is possible to place a crossover line (C)
separating the two regimes.  At high temperature, one recognizes the
same quantum critical phase controlled solely by thermal fluctuation,
while at low temperature, the spectrum of low energy fluctuation in the
gapless magnetic phase controls the NMR relaxation.  When the magnetic
field is raised and crosses $H_{c2}$ (Fig.~\ref{fig15}, right panel),
the same features are qualitatively observed, with possibly a weaker
divergence of $T_1^{-1}$ at low temperature in the gapless phase.  What
is the origin of this critical behavior of $T_1^{-1}$ throughout the
gapless phase?

Because of the broken rotational symmetry, transverse (${\mathcal
S}_\perp$) and longitudinal fluctuations (${\mathcal S}_z$) involve
different processes (cf.~Sec.~6) which do not have the same temperature
dependence.  Since staggered processes, entering ${\mathcal S}_\perp$,
were found to dominate the $T_1^{-1}$ just below $H_{c1}$, it is natural
to first examine the transverse low energy modes in the gapless phase.
In the XXZ mapping for strongly coupled ladders (Sec.~4), two soft modes
(transverse in the valence bond $\sbond, \tbond_{-1}$ basis) were found
(Fig.~\ref{fig5}).  One mode is always at $Q=\pi$, and generates the
staggered process which was already found to be strongly relevant.  The
other mode is at an incommensurate wavevector $Q_\sigma=\pi \sigma$,
which is near $\pi$ when $H$ is close to $H_{c1}$ (top panel,
Fig.~\ref{fig5}).  In \cite{Chitra97}, it was argued that this mode was
gapped and did not contribute to transverse spin-spin correlation, in
apparent contradiction with the spectrum of the XXZ model discussed in
Sec.  4. On the other hand, this incommensurate soft mode in this model
is a transverse singlet-triplet wave.  For physically obscure reasons,
the transverse spin-spin correlator in this state is indeed found to
have zero spectral weight at the incommensurate wavevector.  This point
is crucial since it introduces a fundamental difference between
spin-ladders and integer spin-chains which otherwise belong to the same
universality class.  This has also important consequences for
neutron-scattering studies of spin-ladders.  For NMR, the staggered
process ($Q=\pi$) becomes the only relevant soft mode for (transverse)
relaxation.  If the temperature exceeds the maximum of the lower edge of
the spectrum\footnote{This quantity is proportional the spin-stiffness
of the antiferromagnetic magnons.} between $Q_\sigma=\pi\sigma$ and
$\pi$ (Fig.~\ref{fig5}, top panel), many additional modes contribute to
${\mathcal S}_\perp$.  Hence the $Q=\pi$ soft mode is only relevant for
temperature below the spin-stiffness constant $\rho_s$,
\begin{equation}
\rho_s \equiv \frac{\pi}{2} J_\parallel \frac{M}{M_{sat}} \approx
\sqrt{J_\parallel g \mu_B (H - H_{c1})},
\end{equation}
where the approximation is valid close to $H_{c1}$.
The condition $kT\approx \rho_s$ specifies the crossover line between 
the 'Luttinger liquid' and the quantum critical regime (Fig.~\ref{fig1}
and \ref{fig15} , line C).  As the temperature is lowered, the spectral
weight in the $Q=\pi$ soft mode increases giving the divergent
contribution to
\begin{equation}
\frac{1}{T_1}\bigg\vert_\perp \propto \left(\frac{\rho_s}{\Boltz 
T}\right)^{\eta-1},
\end{equation}
where $\eta$ is the exponent governing the power law decay of the
correlation functions\cite{Chitra97}.  At $H_{c1}$ and $H_{c2}$, the
exponent $\eta$ is known to be $\eta=1/2$ \cite{Sachdev94} and depends
smoothly on the magnetization in between.  The precise dependence of
$\eta$ with $H$ is not known for ladders and could differ from Haldane
systems\cite{Sakai91}, where the incommensurate mode at $Q_\sigma$ is
relevant.  The experimental $T_1^{-1}$ divergence observed
at low temperature appears to be fully consistent with a square-root
singularity (see Fig.~\ref{fig18}).  While there could also be
additional critical fluctuations associated to the 3D
ordering\footnote{Considering that the $Q=\pi$ magnons in the 3D ordered
structure proposed in Sec.~4 are essentially the same modes as in the 1D
quantum disordered phase, the ordering should not have a dramatic effect
on the divergence of the NMR rate.} occurring at very low temperature,
the divergence of $T_1^{-1}$ which is clearly noticeable below $5K$
($\equiv 2J_\parallel$) has to involve 1D fluctuations.

At temperature above $\rho_s$, the relaxation rate gradually crosses 
over to a constant.  This high temperature limit of the relaxation rate
has a maximum around $10\:T$, field at which $g\mu_B H \approx J_\perp$.

\begin{figure}
\centering
\includegraphics[height=50mm, angle=0]{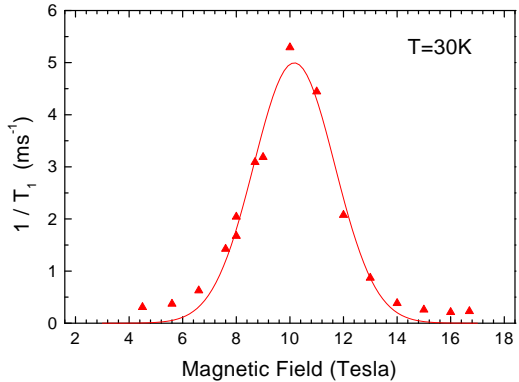}
\caption{Field dependence of the longitudinal relaxation rate 
$T_1{-1}$ as a function of magnetic field.  The 
maximum at $10$T is an independent measure of $J_\perp$.} 
\label{fig16}
\end{figure}
In classical NMR theory\cite{Moriya56}, the high temperature limit of
the $T^{-1}$ relaxation rate is proportional to the second moment of the
spectral distribution of excited states.  This quantity is peaked
precisely in the middle of the triplet band ($m_z=-1$).  At high
temperature where dimers are decorrelated, it is thereore natural to
find a maximum in the relaxation rate as a function of field (shown in
Fig.~\ref{fig16}) at the level crossing between the $\tbond_{-1}$ and
$\sbond$ states.  Since this ''classical'' contribution is proportional
to the zero frequency spectral weight, its temperature dependence is
expected to be weak.  It is anyway unrelated to the hydrodynamic soft
mode at $Q=\pi$, which is a manifestation of the quasi long-range
correlation along the ladder.

The longitudinal fluctuations (${\mathcal S}_z$) have also a
contribution to the $T_1^{-1}$ relaxation rate originating from the
$Q=0$ uniform mode.  They have been shown to be
non-critical\cite{Chitra97}
\begin{equation}
\frac{1}{T_1}\bigg\vert_\parallel \propto \frac{\Boltz T}{\rho_s}.
\end{equation}
Since this contribution is noticeable only close to $H_{c1}$, where it is
weak (see Fig.~\ref{fig15}), it will not be discussed further. 

Since in the experimental data shown in Fig.~\ref{fig15}, we are clearly
able to identify the scaling parameters $x_< = \Delta_-/\Boltz T$ for $H
< H_{c1}$ and $x_> = \rho_s/\Boltz$ for $H>H_{c1}$ (line A 
and C) on each side of the critical field, it is natural to
construct the scaling plots appropriate to this quantum critical point.
\section{Scaling plots in the quantum critical regime}
The concept of scaling at a quantum phase transition, one of the most
beautiful idea in condensed matter physics, was developed in the context
of the metal-insulator transition in disordered
systems\cite{Abrahams79}, where it has been brilliantly applied to doped
semiconductors\cite{Paalanen82}.  But it is in two-dimensions that the
concept has found the most spectacular applications:  disordered
superconducting films \cite{Haviland89,Hebard90} go directly from a
superconducting to an insulating state through a $T=0$ quantum phase
transition as a function of disorder.  Josephson-junction arrays have a
field tune vortex delocalization transition at a critical fraction of
the flux quantum $f_c$\cite{Chen95}.  In a two-dimensional electron gas, 
the transitions between quantum Hall plateaux or to a Hall-insulating
state\cite{Shardar97} are also governed by zero temperature fixed
points\cite{Sondhi97}.  More recently, the scaling properties of a novel
metal-insulator transition\cite{Kravchenko96} in silicon MOSFET's have
also been thoroughly investigated.  In light of this, it is surprising
to find so few experimental studies\cite{Haviland98} of quantum
phase transitions in one dimension.  On the other hand, many 1D systems
have Lorentz invariance:  this confers unique properties to their
quantum critical points.  In particular, their zero temperature critical
behavior can be extended to any temperature via conformal
mapping\cite{Sachdev94}.  This enables to give a precise description of
the finite temperature ''quantum critical regime'' discussed in the
introduction, which is most easily revealed by a scaling analysis.

\begin{figure}
\centering \includegraphics[width=75mm, angle=90]{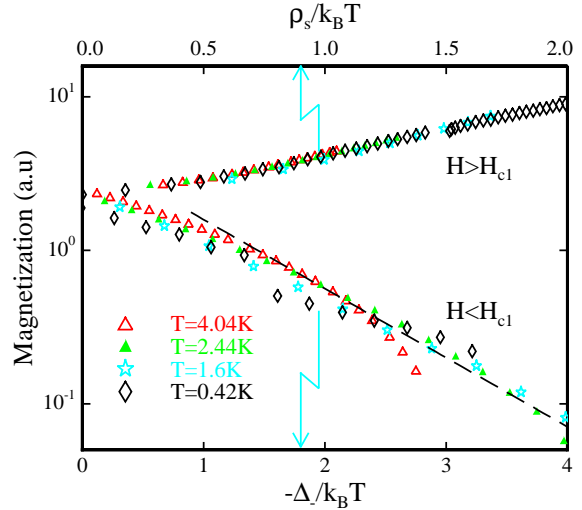}
\caption{Magnetization scaling plot.  Below $H_{c1}$ all curves
collapse on a single curve when plotted in terms of the dimensionless
scaling variable $(\Delta-g\mu_B H)/\Boltz T$.  As expected, the
straight line with slope $-1$ for $x>1$ reproduces the exponential
suppression of the magnetization with an effective gap $\Delta_-$ at low
temperatures.  Above $H_{c1}$, the scaling is also excellent (even in
the ordered phase) using the parameter $\frac{\rho_s}{\Boltz
T}=\frac{\sqrt{g\mu_B(H-H_{c1}) J_\parallel}}{\Boltz T}$.}
\label{fig17}
\end{figure}

In the last section, the scaling parameters $x_> = \rho_s/\Boltz$ 
and $x_< = \Delta_-/\Boltz T$ above and below $H_{c1}$ have been 
identified using NMR.
They are also appropriate to scale the magnetization curves (shown in 
Fig.~\ref{fig7}): the resulting plot is shown in Fig.~\ref{fig17}. 
Considering the overall quality of the scaling, the variable 
$x_<$ and $x_>$ are appropriate to this quantum critical point. 
Furthermore, the $0.42$K curve which crosses into the ordered phase just
above $H_{c1}$ can also be included in this plot.

\begin{figure}
\centering \includegraphics[width=75mm, angle=90]{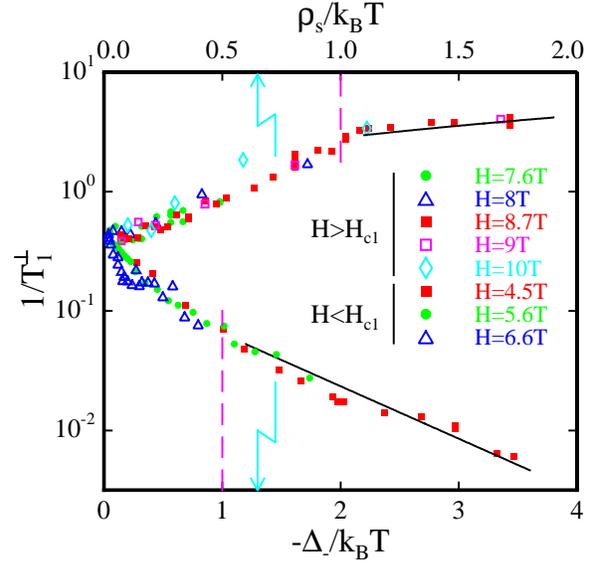}
\caption{Scaling plot.  Below $H_{c1}$ all the $T_1^{-1}$ measurements
collapse on a single curve when plotted in terms of the dimensionless
scaling variable $(\Delta-g\mu_B H)/\Boltz T$.  The point x=1 delimits
the quantum critical regime from the quantum disordered phase.  For
$x>1$, the straight line with slope $-1$ drawn reproduces precisely the
exponential suppression of the $T_1^{-1}$ rate.  Above $H_{c1}$, after 
subtraction of a constant $C(H)$ to the $T_1^{-1}$ rate, it is possible
to scale the remaining critical contributions to the $T_1^{-1}$ 
relaxation in terms of the scaling variable $\frac{\rho_s}{\Boltz 
T}=\frac{\sqrt{g\mu_B(H-H_{c1}) J_\parallel}}{\Boltz T}$.  In the 
Luttinger liquid regime ($x>1$), the data is consistent with 
a scaling function $R(x)=\sqrt{x}$ (solid line).}
\label{fig18}
\end{figure}

A similar scaling analysis of the longitudinal relaxation rate is also
possible.
Below $H_{c1}$, it is straightforward to scale all the experimental
$T_1{-1}$ curves in terms of the single parameter $x_<$, as shown in
Fig.~\ref{fig18} (lower curve).  On this plot, the crossover line (A)
shown on Fig~\ref{fig1} and \ref{fig15} reduces to the single point
$x=1$.  For $x > 1$, the exponentially activated behavior with energy
$\Delta_-$ is the straight line shown with slope $-1$.  Although all
curves scale nicely for all $x$, the quantum critical regime is in
principle limited to $\Boltz T \le 2 J_\parallel$.  Above $H_{c1}$, the
appropriate scaling function is harder to construct because there are
noncritical (classical) contributions to $T_1^{-1}$ which need to be
subtracted.  For all data shown on Fig.~\ref{fig15}, these contributions
were determined so that all curves have, after subtraction, the same
asymptotic limit at high temperature.  The field dependent constant
$C(H)$ substracted was chosen to coincide with the single high-T value
of $T_1^{-1}$ observed in the gapped phase ($H<H_{c1}$, $T\rightarrow
\infty$) and is close to the high temperature limit of $T_1{-1}$ plotted
in Fig.~\ref{fig16}.  In this way the $T\rightarrow \infty$ limit
coincides with the $x_< = x_> = 0$ point of the two scaling functions
(above and below $H_{c1}$).  After subtraction, all data for $H>H_{c1}$
also collapse on a unique scaling curve when expressed in terms of the
scaling parameter $x_>$.  Data points with $x > 1$ are in the Luttinger
liquid regime and a square-root divergence for $T_1^{-1}$ (line shown)
agrees well with the experimental data.  The precise scaling function
can in principle be constructed for all $x$ from the zero temperature
dynamics\cite{Chubukov94} and compared to this experimental scaling
plot.

%
\section{Conclusions}
In this work, several clean experiments on spin-ladders have been used
to illustrate the physics of quantum phase transitions.  Much physical
insight could be drawn using a mapping of strongly coupled ladders on a
much studied model of quantum magnetism, the XXZ model.  For a simple
ladder, the anisotropy $\delta = J_z/J_x =1/2$ puts the system in the
planar (X-Y) universality class between the critical fields $H_{c1}$ and
$H_{c2}$.  Any frustrating coupling (antiferromagnetic cross-bonding
between legs) increases $\delta$.  When $\delta > 1$, the interactions
between fermions (triplets bonds $\tbond_{-1}$) are very large and the
system switches over to an Ising universality class with long range
order.  When an Ising gap is present, plateaux appear in the
magnetization curve in the vicinity of $H_{\rm eff}=0$ ($\Leftrightarrow
H=(H_{c1}+H_{c2})/2$, half filling), as the magnetic field has to
overcome the Ising gap induced by the interactions.  This problem has
also been investigated theoretically by more general
means\cite{Cabra97}.  An interesting possibility raised in this paper is
that similar physics could be induced by much smaller 3D
antiferromagnetic coupling and may already have been observed in
specific heat experiments\cite{Hammar98,Calemzuck98} on CuHpCl.

In the incommensurate gapless phase, low energy dynamical properties are
dominated by the $Q=\pi$ soft mode in the transverse excitation spectrum
(cf.  Fig.~\ref{fig5}).  These fluctuations are not completely quenched
in the gapped phase, since a three-magnons process, which can easily be
identified by the temperature dependence of the NMR relaxation
rate\cite{Chaboussant97b}, remains very strong close to $H_{c1}$
($T_1^{-1}\propto \exp(-2\Delta_-/\Boltz T)$).  At finite temperature,
spin-flip processes introduce a natural cutoff of the spin-spin
correlations.  When $\Boltz T$ is larger than a characteristic quantum
energy (the effective gap $\Delta_-$ below $H_{c1}$ or the
spin-stiffness $\rho_s$ above $H_{c1}$), spin correlation have the same
nature :  this is the quantum critical regime\cite{Chubukov94}.  All
dynamical properties have in this limit a universal behavior which
should be common to quantum phase transition in 1D with a continuous
symmetry.  The scaling properties of the NMR relaxation data on
CuHpCl\cite{Chaboussant98} analyzed in this work should serve as a
reference for future studies of 1D quantum critical points.

In spite of this almost idyllic picture, some
questions remain open.  In the gapless phase, the incommensurate soft
mode should have observable consequences, probably
requiring different experimental probes than the ones considered here.
The nature of the low temperature ordered phase appears to hold the
answer to several key issues in strongly correlated systems.  One
question to be answered is:  can 1D quantum correlation persist in some
form in 3D ordered phases?  Finally, other compounds with more
frustrated magnetic structures can potentially open new horizons in
quantum magnetism:  for example, there are new universality classes
which have not been considered so far.  There are indeed known 
models\cite{Waldtmann98}
exhibiting an energy gap between singlet and triplet sectors but no gaps
in the singlet sector.  The experimental realization of such systems
represents a unique challenge in this field.

The GHMFL is a Laboratoire Conventionn\'e aux Universit\'es J Fourier et
INPG Grenoble I.
\section*{Appendix A: upper critical field}
The energy of the fully polarized state 
$|f\rangle=|\uparrow,\uparrow,\ldots \rangle$ is
\begin{eqnarray}
E_F=\langle F | {\mathcal H} |F \rangle = N \left( \frac{J_\perp}{4} + 
\frac{J_\parallel}{2} -g_z \mu_B H \right)
\nonumber
\end{eqnarray}
The most general state with one unit of angular momentum less than the 
completely polarized state is
\begin{equation}
|\psi \rangle = \sum_{j=1}^{N} \left( u_j S_{2j}^- + v_j S_{2j-1}^-
\right) |F \rangle,
\end{equation}
where the $u_j$ and $v_j$ are respectively the amplitude on the lower 
and upper legs.  They are specified by the condition that $\psi \rangle$ 
should be an eigenstate of $\mathcal H$ with energy $E_-$, or 
equivalently
\begin{equation}
E_F - E_- = \sum_{j=1}^{N} \left[u_j S_{2j}^- + v_j S_{2j-1}^- , 
{\mathcal H} \right] |F \rangle.
\end{equation}
This condition yields a set of two coupled equations for the $u_j$ and 
$v_j$, which are easily solved in Fourier space.  The dispersion 
relations for the corresponding spin-wave modes 
$\epsilon^\alpha (k)=E_F-E^\alpha_-(k)$ are \begin{eqnarray}  
\epsilon^o(k) & = & g_z \mu_B H - J_\parallel (1-\cos k), \label{eqA1}
\\
\epsilon^a(k) & = & g_z \mu_B H - J_\perp - J_\parallel (1-\cos k). 
\label{eqA2}
\end{eqnarray}
In the strong coupling limit, the 'acoustic' mode (\ref{eqA2}) becomes
soft first at wavevector $k=\pi$, when the field drops below the field
which specifies $H_{c2}$ (\ref{Hc2}).  It is straightforward to generalize 
the argument to more complex systems.
\section*{Appendix B: thermodynamics of the XXZ model}
This problem has been completely set out by Takahashi and 
Suzuki\cite{Takahashi72}.  For an anisotropy factor of $J_z/J_x=1/2$,
all thermodynamic quantities can be computed from the solution of the 
coupled set of integral equations for the functions $\eta(x)$ and 
$\kappa(x)$,
\begin{eqnarray}
\ln \eta(x) & = & 3\sqrt{3}\frac{J_\parallel}{\Boltz T} + s(x) * 
\ln(1+u(x)) \nonumber\\
u(x) & = & 2 \kappa (x) \cosh \frac{3g\mu_B H_{eff}}{2 \Boltz T} + 
\kappa^2 (x) \\
\ln \kappa (x) & = & s(x) * \ln (1+ \eta(x)) \label{TS1}
 \end{eqnarray}
where $s(x) = \frac{1}{4} {\rm sech} \frac{\pi x}{2}$ and $*$ is the 
convolution product of two functions.  For each value of the 
temperature and of the effective field, the thermodynamics is 
specified by the value $\kappa (0)$, i.e. the free-energy per spin is
\begin{equation}
\frac{F}{N} = -\frac{J_{\parallel}}{4} -\Boltz T \ln \kappa (0).
\end{equation}
We solved Eqs. (\ref{TS1}) iteratively from the known 
solutions, $\eta(x)=3$ and $\kappa(x) =2$ for $J_\parallel=0$ and 
$H_{eff}=0$.  We checked our results against the power series 
expansion in $J/T$\cite{Takahashi72}
\begin{eqnarray}
\frac{F}{N} & = &  -\Boltz T \ln \left( 2 \cosh \frac{g\mu H}{2\Boltz
T}\right) - \frac{J}{8} \frac{1}{\cosh^2 \left( \frac{g\mu H}{2\Boltz T}
\right)} \nonumber \\
& - &\frac{3 J^2}{32 \Boltz T} \left\{ \frac{1}{\cosh^2 \left(
\frac{g\mu H}{2\Boltz T} \right)} -\frac{1/4}{\cosh^4 \left(
\frac{g\mu H}{2\Boltz T} \right)} \right\} \nonumber
\end{eqnarray}
The comparison with the experimental results shown in Fig.~\ref{fig7} 
is really excellent.


\begin{thebibliography}{9}
\bibitem{hightemp} Since all the high-energy states are excluded in the 
XXZ mapping, its validity is limited to temperatures below the 
$J_\perp/\Boltz$  .Similarly, the Zeeman splitting must be sufficiently 
large (say $g\mu_B H > J_\perp/2$) to avoid admixture with higher energy
states.
\end{thebibliography}
 \end{document}